\documentclass[10pt, conference]{IEEEtran}
\usepackage{paralist}
\usepackage{graphicx}

\usepackage{filecontents}

\ifCLASSINFOpdf
\else
\fi
\begin{document}
%
\title{
Towards the Assessment of Stress and Emotional Responses of a Salutogenesis-Enhanced Software Tool Using Psychophysiological Measurements }


\author{\IEEEauthorblockN{Jan-Peter Ostberg, Daniel Graziotin, Stefan Wagner}
\IEEEauthorblockA{Institute for Software Technology, University of Stuttgart, Germany\\
Email: \textit{firstName}.\textit{lastName}@informatik.uni-stuttgart.de}
\IEEEauthorblockN{Birgit Derntl}
\IEEEauthorblockA{Department of Psychiatry and Psychotherapy, University of T\"ubingen, Germany\\
Email: birgit.derntl@med.uni-tuebingen.de}

}

%


\maketitle

\begin{abstract}
Software development is intellectual, based on collaboration, and performed in a highly demanding economic market. As such, it is dominated by time pressure, stress, and emotional trauma.  While studies of affect are emerging and rising in software engineering research, stress has yet to find its place in the literature despite that it is highly related to affect. In this paper, we study stress coping with the affect-laden framework of Salutogenesis, which is a validated psychological framework for enhancing mental health through a feeling of coherence. We propose a controlled experiment for testing our hypotheses that a static analysis  tool enhanced with the Salutogenesis model will bring 
\begin{inparaenum}
\item a higher number of fixed quality issues,
\item reduced cognitive load,
\item reduction of the overall stress, and
\item positive affect induction effects to developers.
\end{inparaenum} The experiment will make use of validated physiological measurements of stress as proxied by cortisol and alpha-amylase levels in saliva samples, a psychometrically validated measurement of mood and affect disposition, and stress inductors such as a cognitive load task. Our hypotheses, if empirically supported, will lead to the creation of environments, methods, and tools that alleviate stress among developers while enhancing affect on the job and task performance.
\end{abstract}


%
\IEEEpeerreviewmaketitle

\section{Introduction}

The development of software systems is a series of complex intellectual activities involving the collaboration of individuals. The resulting artifacts are intangible, and this poses challenges with respect to the perception of project progress from the point of view of developers, managers, and  customers. The current economic systems demand short time to market, high quality, and the necessity to stay in budget. This complex intertwining of factors and demands is ruled by time pressure, stress, and emotional trauma \cite{Wastell1993,graziotin2015you,Graziotin2014IEEESW,Mantyla:2014:TPC:2568225.2568245, rajeswari2003development}. 

Stress is the scourge of the modern industrial world \cite{de1998psychosocial} and mental-health diseases related to the working environment are on the rise \cite{lademann2006psychische}. While studies of affect are emerging and rising in software engineering research, stress has yet to find its place in the software engineering literature, with very few studies scattered through the last two decades. While these studies are desirable, they are based on the survey approach.

  In this paper, we are opening a line of research of stress in software engineering, and we start with a study of stress coping with the affect-laden framework of Salutogenesis~\cite{antonovsky1979health}. Salutogenesis (detailed in the next section) is a validated method which supports mental health through a feeling of coherence. The method can be applied to software development processes, methods, and tools.

We are presenting our research proposal for a controlled experiment of stress coping mechanisms in software quality enhancements. We hypothesize that a software tool (FindBugs), when enhanced using the Salutogenesis model, will bring 
\begin{inparaenum}
\item a higher number of fixed quality issues,
\item higher cognitive capacity,
\item reduction of the overall stress, and
\item positive affect induction effects
\end{inparaenum}
compared to the non-enhanced version of the tool. The controlled experiment will make use of validated physiological measurements of stress as proxied by cortisol and $\alpha$-amylase levels in saliva samples~\cite{noto2005relationship}, a psychometrically validated measurement of mood and affect disposition \cite{watson1988development}, and stress inductors such as a cognitive load task test~\cite{gevins1993neuroelectric}. This will enable us to test out causality chains and indicators from the endocrine system, which will bring us to a stronger empirical case. If our hypotheses are empirically validated, the results will offer a new method for enhancing the quality of software systems while reducing the overall stress of developers.


In the rest of the present paper, we will talk about stress and affect theory, and then we will explain our proposed experiment design. Because of space limitations, we cite related work where fitting.

\section{Stress and Salutogenesis}

\subsection{Stress, Affect, and Software Development}
We have adopted Weinert's~\cite{weinert2004b} definition of stress as ``\ldots an adaptive reaction to exceeding psychic or physical demands of the surroundings.''. Constraints and demands are connected to the build up of stress \cite{weinert2004b,zapf2004stress}. The following conditions have to be met for the generation of a stress response by the individual. 

\begin{inparaenum}[(c1)]
\item{\emph{The outcomes of the triggering event must not be known beforehand.}}
\item{\emph{The results of the trigger have to be important to the individual.} }
\end{inparaenum}
 For example, stress arises if an individual cannot be sure that a deadline can be met and that the consequences could be catastrophic, e.g. loss of employment.

One of the most prominent stress process theories, namely the General Adaptation Syndrome (GAS) by Selye~\cite{selye1946general}, has built upon the previously defined factors and conditions. GAS defines three stages, which we outline together with a software engineering scenario.
\begin{inparaenum}
\item{\emph{The Alarm Phase:} The individual recognizes the triggering event and reacts to it, according to the severity of the event. In extreme cases, a blockade can happen, which hinders the individual to act. For example, a developer could enter the alarm phase after a sudden report of hundred problems by a static analysis tool. The event would not be expected, and an immediate solution to the problem would have yet to emerge.}
\item{\emph{The Resistance Phase:} The individual tries to antagonize the stress triggering event and overcome it. If the individual succeeds in overcoming the triggering event, the stress situation ends at this phase. In our scenario, the developer would now start the triage process and fixing the issues, one by one.}
\item{\emph{The Exhaustion Phase:} If the stress generating situation persist, it might lead to psychophysiological exhaustion. The individual stops resisting and slips into a form of blockade. The blockade can range from minor physical problems to paralysis (like shell-shocked soldiers in the first world war) and depressive states even leading to suicidal tendencies \cite{kaufmann1982arbeitsbelastung} \cite{rensing2006mensch} \cite{chen2009work}. For our unlucky developer, the exhaustion phase might start when they run out of ideas for solving the problems at hand, and no exiting solutions have emerged yet. The developer would now stop resisting and surrender to the impeding situation. The consequences of the exhaustion phase, unless mediated by changing task, obtaining help, or fixing the issues, could lead to serious health issues reaching from rather harmless psychosomatic back pain to burnout or depression or even death by an heart attack \cite{litzcke2010stress}.}
\end{inparaenum}

Our scenario will likely elicit an empathic response to the reader, as it \emph{looks like a negative experience}. Negative experiences are dominated by negative affect \cite{Diener2009}. A link between stress and affect seems intuitive. Feldman et al. \cite{Feldman:1999cz} have reported that traditional views of stress and disease suggest that our appraisal of threats with negative emotions leads to physiological changes that influence disease onset and progression. Indeed, the relationship between stress and affect is complex as the two constructs are intertwined. In a study of 420 individuals, stress was found positively correlated with the state affect of anxiety and depression \cite{Spada:2008hc}. Physiological responses such as cardiovascular ones were also found to be positively correlated with negative emotions and stressor tasks \cite{Feldman:1999cz}. Furthermore, workplace settings require emotional regulation in response to stress, which causes emotional dissonance that is linked to detriment of psychological health \cite{Gross:1998fp,Zapf:2001dna}. Although the cause-and-consequence relationship type has yet to be explored, the correlation between affect, stress, and negative physiological responses has been established. Software developers are not immune to the stress threat, as the development of software is a very stressful activity \cite{Wastell1993}. We found out recently that stress is an output of unhappiness and negative affect among software developers \cite{Graziotin:2017}, and it is known that stress leads to burnout among developers \cite{sonnentag1994stressor}. Yet, mediating the negative affect of developers has been theorized to be effective in terms of unhappiness mediation and the subsequent programming performance boost \cite{graziotin2015you}.

Summarizing, stress is bad for all workers, developers included (if not even more damaging than with other jobs), and its reduction should be of importance to researchers of software engineering and practitioners. The Salutogenesis model is a proposal to counteract stress, and it can be integrated into the software development process as well as development tools.

\subsection{Salutogenesis: an Affect-laden Framework to Reduce Stress}
Salutogenesis was introduced by Antonovsky~\cite{antonovsky1979health} and later recalled by a review study by Eriksson and Lindstr\"om~\cite{lindstrom2005Salutogenesis}. Salutogenesis is a model to explain the origin of health as in contrast to pathogenesis, which explores instead the manner of development of a disease. The model considers the concept of \emph{health} as a continuum between total ill health and total health. The health continuum is influenced by affect (as companions of stressors and resistors), and attitudes. An important influencer is the feeling of coherence, which is a feeling of being consistent with the happenings in our life, which fosters the emergence and steadiness of health. Salutogenesis identifies three pillars that support health by boosting the feeling of coherence.

\begin{inparaenum}
\item{\emph{Understandability}, or comprehensibility, enables humans to recognize the known and unknown stimuli of life as ordered and structured events and not as chaotic or arbitrary. For example, well commented and structured code brings a sense of understandability of the system.}
\item{\emph{Manageability}, or the ability to overcome obstacles, means the degree to which someone has the resources to overcome a given tasks. For example, having the time to do a job right rather then having to cut work short or leaving parts unfinished.}
\item{\emph{Meaningfulness}, or the degree of sense of a stimulus. It reflects how much a situation or task makes sense to individuals and how important the stimulus is perceived. Meaningfulness is also an important influencer of motivation, and it adds to seeing a task as a challenge rather then a burden. For example, showing workers how their task's output contributes to the success of a whole project rather then just telling them to just do their work.}
\end{inparaenum}

The extent to which these three factors have to be met for the individual to stay healthy, experiencing less stress, depends on the individual's general resistance resources~\cite{antonovsky1979health}. The amount and types of these resources which are available to the individual depends on genetic, constitution and psychosocial aspects like intelligence, commitment, general attitude towards life, and basic mood~\cite{antonovsky1979health}.

Salutogenesis predicts that the more these three values apply, the deeper the feeling of coherence is experienced and so living is experienced as less stressful, reducing the likelihood of becoming ill. Studies have found support for the model's claims~\cite{lindstrom2005Salutogenesis} \cite{eriksson2006antonovsky} \cite{eriksson2005validity}. 

\section{FindBugs and Enhancements}
FindBugs is a static analysis tool. It was born out of the idea to catch trivial mistakes developers make on a daily basis--which FindBugs calls \textit{bug patterns}.
We made enhancements to the Eclipse plugin of FindBugs based on the three pillars of  Salutogenesis~\cite{antonovsky1979health}. Because of space restrictions, we enlist here only one enhancement per pillar. We reported all enhancements elsewhere~\cite{ostberg2016ease}. 

We introduced the ability to leave comments to a class of bug patterns to communicate thoughts on a bug, e.g., why it is a false positive or why it is highly important to fix that one. This increases the \textit{understandability}. 

To increase the \textit{manageability}, we created buttons that will reduce the amount of warnings shown in five levels to offer the users their preferred level of manageability.

\textit{Meaningfulness} is increased by the possibility to mark false positive warnings which are first sorted into a special section of FindBugs' \emph{bug explorer} and are hidden after a new analysis.

\section{Experiment Design}
\begin{figure*}[tb]
\centering
\includegraphics[width=0.9\textwidth]{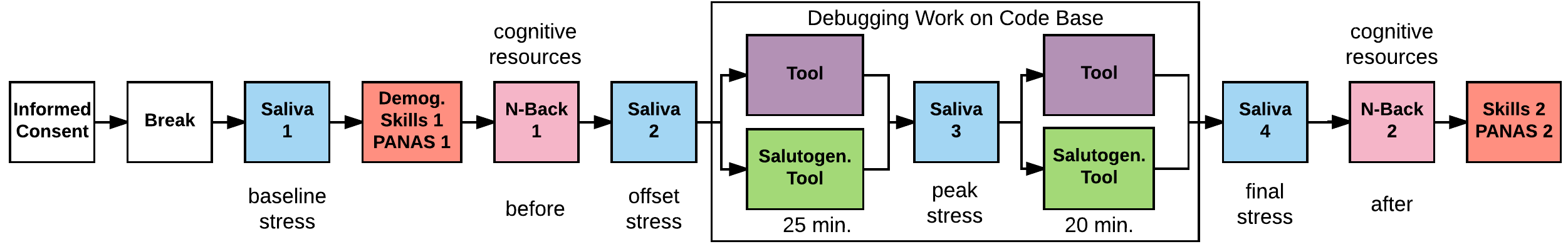}
\caption{Steps of the Proposed Experiment for Assessing Stress and Emotional Responses of a Salutogenesis-Enhanced Software Tool}
\label{fig:phases}
\end{figure*}

To test our hypothesized improvements (higher number of fixed issues, higher cognitive capacity, reduction of the overall stress, positive affect induction effects) of applying Salutogenesis model to the static analysis tool FindBugs, we propose the following controlled experiment, which we represent in Fig. \ref{fig:phases}.

After expressing written informed consent, the participants will have 5 minutes to rest. This is one of our strategies for limiting endogenous stressors that could originate before starting the experiment.

We will take a first saliva sample (azure boxes in Fig. \ref{fig:phases}) to establish the \emph{baseline stress} measures for the cortisol and $\alpha$-amylase measurements. Cortisol and $\alpha$-amylase are two stress indicators, well established in the medical stress research~\cite{noto2005relationship}. 

To assess potential stress sources, the participants will then fill in questionnaires (red boxes in Fig. \ref{fig:phases}): Demographic questions such as gender, age, pre-existing neurological or psychological conditions, and medical drug intake will let us control the hormone level measurement. To keep a low noise within the measurements, we ask the participants to not consume beverage containing sugar (including unsweetened tea or coffee) or smoke one hour before the experiment. We will add items related to stress perception and three factors to assess the participants' debugging skills and the participants' self evaluation of these skills. The self evaluation is an indicator for the coherence feeling. A third questionnaire aims at self-efficacy which gives us some hints on the stress resilience of the participants. We use the PANAS scale~\cite{watson1988development} to assess the participants' mood.

The next step measures the participants cognitive capacity (rose boxes), as our hypotheses imply that stress reduced working as proxied by Salutogenesis frees cognitive capacity. We will use the N-Back-Test to assess the cognitive capacity~\cite{gevins1993neuroelectric} which is implemented and run in the PEBL environment~\cite{mueller2014psychology}.

We will take a second saliva sample to be able to assess the stress generated by these questionnaires and the cognitive load test (\emph{offset stress}), so that we can distinguish between the build up of stress through the cognitive testing and the stress generated in the debugging task.

The participants are randomly assigned to two balanced groups: those using the vanilla FindBugs (violet boxes) and those using the Salutogenesis-enhanced FindBugs (green boxes). Each participant will work on the same codebase (the open-source project \textit{Sweet Home 3D}). A FindBugs analysis on the code base reveals about 100 findings covering every severity and confidence rank as well as over 20 different categories of findings.

The participants will use the tool for 25 min., after which we take a third saliva sample which is expected to be a \emph{peak stress} moment according to the literature. 

After that, the participants will work to the code for other 20 min.\ The task ends with a fourth and last saliva sample, which represents the \emph{final stress} measure. 

To assess the cognitive resources consumed by the previous work on the code we have the participants take the N-Back-Test again. To lessen any possible learning effects we will use a different N-Back style for the second measurement.

Finally, to assess the potential influence of the Salutogenesis-enhanced version of FindBugs, we ask the participants to fill out a second PANAS and knowledge questionnaires.
 
 As stress inductors, besides the debugging task itself, we will set up a contest where those that are able to fix at least 6 findings from 4 different categories can win a 10,00 Euro voucher for the university cafeteria.  Also, participants will be aware that a public (yet anonymized) ranking of the correctly fixed FindBugs findings will be published.

The results of the work on the code with both FindBugs versions will be reviewed to count the successfully fixed findings. Screen captures will be checked for potential anomalies and for assessing  solution strategies by the participants. The count of the successfully fixed findings will be used to determine the winners of the voucher.

The experiment procedure will be implemented by following the Helsinki declaration's~\cite{helsinki2013} ethical and privacy considerations. We are already collaborating with the state's data protection agency (Zendas) in regards.

%
%

\section{Conclusion and Future Work}
In this paper, we have proposed a controlled experiment for testing our hypotheses that a static analysis tool enhanced with Salutogenesis will bring 
\begin{inparaenum}
\item a higher number of fixed software quality issues,
\item reduced cognitive load,
\item reduction of the overall stress, and
\item positive affect induction effects to developers\end{inparaenum}.

Compared to other ways to evaluate stress---such as the measurement of skin resistance, heart rate, or adrenaline in blood---our method does not require on-site support of medically trained staff, is easily applicable on larger groups, and is more objective than textual psychological tests (e.g., questionnaires). While the assessment of the results still requires the involvement of a chemistry laboratory, just a couple of researchers would be able to gather hundreds of samples. Yet, our proposed design is complex, and its execution requires significant effort. We aim to pilot the experiment before attending SEmotion'17. We are still looking for improvements to the design, measurement techniques, and reducing costs and effort.

Shall our hypotheses find empirical support, our study results and our proposed improvements based on the Salutogenesis model have the potential to offer a new method for enhancing the quality of software systems while reducing the overall stress of developers.

\section*{Acknowledgment}
Daniel Graziotin has been supported by the Alexander von Humboldt (AvH) Foundation.



%
\bibliographystyle{IEEEtran}
\bibliography{Literatur}

\begin{thebibliography}{10}
\providecommand{\url}[1]{#1}
\csname url@samestyle\endcsname
\providecommand{\newblock}{\relax}
\providecommand{\bibinfo}[2]{#2}
\providecommand{\BIBentrySTDinterwordspacing}{\spaceskip=0pt\relax}
\providecommand{\BIBentryALTinterwordstretchfactor}{4}
\providecommand{\BIBentryALTinterwordspacing}{\spaceskip=\fontdimen2\font plus
\BIBentryALTinterwordstretchfactor\fontdimen3\font minus
  \fontdimen4\font\relax}
\providecommand{\BIBforeignlanguage}[2]{{%
\expandafter\ifx\csname l@#1\endcsname\relax
\typeout{** WARNING: IEEEtran.bst: No hyphenation pattern has been}%
\typeout{** loaded for the language `#1'. Using the pattern for}%
\typeout{** the default language instead.}%
\else
\language=\csname l@#1\endcsname
\fi
#2}}
\providecommand{\BIBdecl}{\relax}
\BIBdecl

\bibitem{Wastell1993}
D.~Wastell and M.~Newman, ``{The behavioral dynamics of information system
  development: A stress perspective},'' \emph{Accounting, Management and
  Information Technologies}, no.~2, pp. 121--148, 1993.

\bibitem{graziotin2015you}
D.~Graziotin, X.~Wang, and P.~Abrahamsson, ``{How do you feel, developer? An
  explanatory theory of the impact of affects on programming performance},''
  \emph{PeerJ Computer Science}, vol.~1, no.~1, p. e18, Aug. 2015.

\bibitem{Graziotin2014IEEESW}
------, ``{Software Developers, Moods, Emotions, and Performance.}'' \emph{IEEE
  Software}, vol.~31, no.~4, pp. 24--27, 2014.

\bibitem{Mantyla:2014:TPC:2568225.2568245}
M.~V. M\"{a}ntyl\"{a}, K.~Petersen, T.~O.~A. Lehtinen, and C.~Lassenius, ``Time
  pressure: A controlled experiment of test case development and requirements
  review,'' in \emph{Proc.~36th International Conference on Software
  Engineering (ICSE)}, 2014, pp. 83--94.

\bibitem{rajeswari2003development}
K.~Rajeswari and R.~Anantharaman, ``Development of an instrument to measure
  stress among software professionals: Factor analytic study,'' in
  \emph{Proc.~2003 SIGMIS Conference on Computer Personnel Research}.\hskip 1em
  plus 0.5em minus 0.4em\relax ACM, 2003, pp. 34--43.

\bibitem{de1998psychosocial}
B.~P. de~Jonge, B.~Jan, J.~F. Ybema, and C.~J. de~Wolff, ``Psychosocial aspects
  of occupational stress,'' \emph{Handbook of work and organizational
  psychology: Work psychology}, vol.~2, p. 145, 1998.

\bibitem{lademann2006psychische}
J.~Lademann, H.~Mertesacker, and B.~Gebhardt, ``{Psychische Erkrankungen im
  Fokus der Gesundheitsreporte der Krankenkassen},''
  \emph{Psychotherapeutenjournal}, vol.~2, no. 2006, pp. 123--139, 2006.

\bibitem{antonovsky1979health}
A.~Antonovsky, ``Health, stress and coping: New perpectives on mental and
  physical well-being,'' \emph{YosseyBass}, 1979.

\bibitem{noto2005relationship}
Y.~Noto, T.~Sato, M.~Kudo, K.~Kurata, and K.~Hirota, ``The relationship between
  salivary biomarkers and state-trait anxiety inventory score under mental
  arithmetic stress: a pilot study.'' \emph{Anesthesia \& Analgesia}, vol. 101,
  no.~6, pp. 1873--1876, 2005.

\bibitem{watson1988development}
D.~Watson, L.~A. Clark, and A.~Tellegen, ``Development and validation of brief
  measures of positive and negative affect: the panas scales.'' \emph{Journal
  of personality and social psychology}, vol.~54, no.~6, p. 1063, 1988.

\bibitem{gevins1993neuroelectric}
A.~Gevins and B.~Cutillo, ``Neuroelectric evidence for distributed processing
  in human working memory,'' \emph{Electroencephalography and Clinical
  Neurophysiology}, vol.~87, pp. 128--143, 1993.

\bibitem{weinert2004b}
A.~Weinert, \emph{{Organisations- und Personalpsychologie}}.\hskip 1em plus
  0.5em minus 0.4em\relax Beltz PVU, 2004.

\bibitem{zapf2004stress}
D.~Zapf and N.~K. Semmer, ``{Stress und Gesundheit in Organisationen},''
  \emph{Enzyklop{\"a}die der Psychologie, Themenbereich D, Serie III}, vol.~3,
  pp. 1007--1112, 2004.

\bibitem{selye1946general}
H.~Selye, ``The general adaptation syndrome and the diseases of adaptation 1,''
  \emph{The Journal of clinical endocrinology \& metabolism}, vol.~6, no.~2,
  pp. 117--230, 1946.

\bibitem{kaufmann1982arbeitsbelastung}
I.~Kaufmann, H.~Pornschlegel, and I.~Udris, ``{Arbeitsbelastung und
  Beanspruchung},'' \emph{Humane Arbeit--Leitfaden f{\"u}r Arbeitnehmer},
  vol.~5, pp. 13--48, 1982.

\bibitem{rensing2006mensch}
L.~Rensing, \emph{Mensch im Stress: Psyche, K{\"o}rper, Molek{\"u}le}.\hskip
  1em plus 0.5em minus 0.4em\relax Elsevier, Spektrum Akad. Verlag, 2006.

\bibitem{chen2009work}
W.-Q. Chen, O.-L. Siu, J.-F. Lu, C.~L. Cooper, and D.~R. Phillips, ``Work
  stress and depression:the direct and moderating effects of informal social
  support and coping,'' \emph{Stress and Health}, vol.~25, pp. 431--443, 2009.

\bibitem{litzcke2010stress}
S.~M. Litzcke and H.~Schuh, \emph{Stress, Mobbing und Burn-out am
  Arbeitsplatz}.\hskip 1em plus 0.5em minus 0.4em\relax Springer-Verlag, 2010,
  ch. 1.6, pp. 29--35.

\bibitem{Diener2009}
E.~Diener, D.~Wirtz, W.~Tov, C.~Kim-Prieto, D.~Choi, S.~Oishi, and
  R.~Biswas-Diener, ``{New Well-being Measures: Short Scales to Assess
  Flourishing and Positive and Negative Feelings},'' \emph{Social Indicators
  Research}, vol.~97, no.~2, pp. 143--156, 2009.

\bibitem{Feldman:1999cz}
P.~J. Feldman, S.~Cohen, S.~J. Lepore, K.~A. Matthews, T.~W. Kamarck, and A.~L.
  Marsland, ``Negative emotions and acute physiological responses to stress,''
  \emph{Annals of Behavioral Medicine}, vol.~21, pp. 216--222, 1999.

\bibitem{Spada:2008hc}
M.~M. Spada, A.~V. Nik{\v c}evi{\'c}, G.~B. Moneta, and A.~Wells,
  ``{Metacognition, perceived stress, and negative emotion},''
  \emph{Personality and Individual Differences}, vol.~44, no.~5, pp.
  1172--1181, Apr. 2008.

\bibitem{Gross:1998fp}
J.~J. Gross, ``{Antecedent-and response-focused emotion regulation: divergent
  consequences for experience, expression, and physiology.}'' \emph{Journal of
  Personality and Social Psychology}, vol.~74, no.~1, pp. 224--237, 1998.

\bibitem{Zapf:2001dna}
D.~Zapf, C.~Seifert, B.~Schmutte, H.~Mertini, and M.~Holz, ``{Emotion work and
  job stressors and their effects on burnout},'' \emph{Psychology {\&} Health},
  vol.~16, no.~5, pp. 527--545, 2001.

\bibitem{Graziotin:2017}
D.~Graziotin, F.~Fagerholm, X.~Wang, and P.~Abrahamsson, ``Unhappy developers:
  Bad for themselves, bad for process, and bad for software product,'' in
  \emph{Proc.\ 2017 IEEE/ACM 39th International Conference on Software
  Engineering Companion (ICSE-C)}.\hskip 1em plus 0.5em minus 0.4em\relax ACM,
  2017.

\bibitem{sonnentag1994stressor}
S.~Sonnentag, F.~C. Brodbeck, T.~Heinbokel, and W.~Stolte, ``Stressor-burnout
  relationship in software development teams,'' \emph{Journal of occupational
  and organizational psychology}, vol.~67, pp. 327--341, 1994.

\bibitem{lindstrom2005Salutogenesis}
B.~Lindstr{\"o}m and M.~Eriksson, ``Salutogenesis,'' \emph{Journal of
  Epidemiology and community health}, vol.~59, no.~6, pp. 440--442, 2005.

\bibitem{eriksson2006antonovsky}
M.~Eriksson and B.~Lindstr{\"o}m, ``Antonovsky's sense of coherence scale and
  the relation with health: a systematic review,'' \emph{Journal of
  epidemiology and community health}, vol.~60, no.~5, pp. 376--381, 2006.

\bibitem{eriksson2005validity}
------, ``Validity of antonovsky's sense of coherence scale: a systematic
  review,'' \emph{Journal of epidemiology and community health}, vol.~59,
  no.~6, pp. 460--466, 2005.

\bibitem{ostberg2016ease}
J.-P. Ostberg and S.~Wagner, ``At ease with your warnings: the principles of
  the salutogenesis model applied to automatic static analysis,'' in
  \emph{Proc.\ IEEE 23rd International Conference on Software Analysis,
  Evolution, and Reengineering (SANER)}.\hskip 1em plus 0.5em minus 0.4em\relax
  IEEE, 2016, pp. 629--633.

\bibitem{mueller2014psychology}
S.~T. Mueller and B.~J. Piper, ``The psychology experiment building language
  ({PEBL}) and {PEBL} test battery,'' \emph{Journal of neuroscience methods},
  vol. 222, pp. 250--259, 2014.

\bibitem{helsinki2013}
{World Medical Association}, ``World medical association declaration of
  {Helsinki}: Ethical principles for medical research involving human
  subjects,'' \emph{JAMA}, vol. 310, no.~20, pp. 2191--2194, 2013.

\end{thebibliography}

\end{document}